\documentclass[11pt]{article} \usepackage{moriond2000,epsfig}

\def\be{\begin{equation}}
\def\ee{\end{equation}}
\def\bea{\begin{eqnarray}}
\def\eea{\end{eqnarray}}

\def\dlsodos^#1 {{D_{\rm LS}^{#1}\over D_{\rm OS}^{#1}}}
\def\mA{{\cal A}}

\def\mg{\big<}
\def\md{\big>}
\def\be{\begin{equation}}
\def\ee{\end{equation}}
\def\ba{\begin{eqnarray}}
\def\ea{\end{eqnarray}}
\def\d{{\rm d}}
\def\xs{x_{\rm str.}}
\def\ys{y_{\rm str.}}

\def\ncrit{n_{\rm crit.}(x)}
\def\nbcrit{\overline{n}_{\rm crit.}(x)}

\def\nb{\overline{n}}
\def\critb{{d_{\rm crit.}}}
\def\lcrit{L_{\rm crit.}}

\begin{document}
\vspace*{4cm}
\title{COSMIC STRINGS LENS PHENOMENOLOGY REVISITED}

\author{ Francis BERNARDEAU$^1$ \& Jean-Philippe UZAN$^2$}

\address{$^1$Service de Physique Th\'eorique, CE de Saclay\\
         F-91191 Gif-sur-Yvette Cedex, France\\
         $^2$Laboratoire de Physique Th\'eorique, UMR-8627 du CNRS,
         universit\'e Paris XI, \\F-91405 Orsay Cedex, France
}

\maketitle  

\abstracts{ We present investigations of lens phenomenological properties of
cosmic strings for deep galaxy  surveys.  General results that have obtained
for  lineic  energy  distribution  are  presented  first.   We  stress  that
generically  the local convergence  always vanishes  in presence  of strings
although  there might  be some  significant distortions.  We then  propose a
simplified  model  of  strings,  we  call  ``Poisson  strings'',  for  which
exhaustive investigations can be done either numerically or analytically.}

\section{Motivations}

At a  time when the  observational data seem  to converge towards  models of
structure  formation  with  scalar  adiabatic initial  fluctuations  obeying
Gaussian statistics~\cite{Boom}, it is  probably worth recalling that cosmic
strings, and  more generally tolopological defects, form  under very general
circumstances.  Therefore,  although they may not  be the main  seeds of the
large-scale structure of the universe  or of the CMB anisotropies, relics of
such objects due to early time  phase transitions may still exist.  With the
advance of  a new generation  of telescopes and  large field CCD  cameras we
think it is worth to keep in mind what could be their observational effects.

In this presentation  we are thus interested in  the observational signature
of cosmic  strings on background  galaxies. Future large-scale  surveys that
are in preparation for weak lensing observations and that are going to cover
a fair  fraction of the sky  with unprecedented image quality  are a natural
playground for elaborating detection strategies of such strings.


\section{General lens properties of strings}

\subsection{Uniform straight strings}

The case of uniform straight  string (with Goto-Nambu equation of state) has
been exhaustively  described in the literature~\cite{VS}.  In  this case the
metric is actually flat around the  string. The only effect of the string is
to induce a ``missing angle'' so that space is conical around the string. As
a result galaxies that are behind  the string may exhibit double images. The
pair  separation of  these images  is  directly proportional  to the  lineic
energy  density of string  $\mu_0$.  For  strings that  formed at  the Grand
Unification  scale, that  would corresponds  to  image separation  of a  few
arcsecs. Optical investigations, for that respect, seem the most appropriate
way of detecting such strings.

However, the  idea strings  induce series of  image pairs is  however coming
from  a simplified  description  of the  string  energy properties.   Simple
numerical experiments~\cite{dLKV} done  with numerical simulations of string
networks  suggest  that  this  is  too  naive  a  view  and  that  the  lens
phenomenology of strings is much more complex.

\subsection{General string effects}

\begin{figure}
\centering {\resizebox*{7.5cm}{!}{\includegraphics{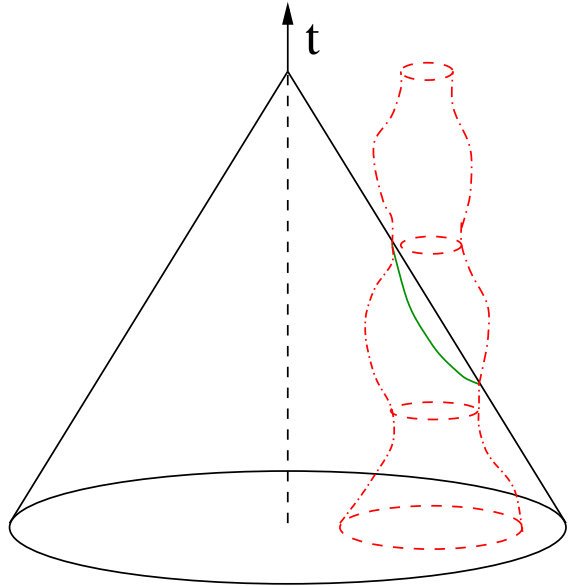}}}
\caption
{Schematic  view of the  source term  for the  lens effect  of a  string. It
corresponds to the  intersection of the loop world-sheet  and the past light
cone of the observer.}
\label{intersec}
\end{figure}

In general  the displacement  field induced by  strings is not  uniform, and
thus significant deformation effects  of background galaxies can be induced.
One can show that, under very general hypothesis (with respect to the string
equation of  state and  geometry), there exists  a potential from  which the
elements  of the deformation  matrix derive~\cite{UB}.   It can  formally be
written in terms of the angular positions $(x,y)$,
\begin{equation} 
\varphi(x,y)=4\,G\,\dlsodos^{   }  \int\d   s\
\mu[\xs(s),\ys(s)]\ \log\left([x-\xs(s)]^2+[y-\ys(s)]^2\right)^{1/2}         
\end{equation}        
where $(\xs(s),\ys(s))$  are the angular string coordinates  for the angular
curvilinear  position $s$,  $\mu(\xs(s),\ys(s))$ is  the  ``projected energy
density''  at those  positions,  $G$  is the  Newton  constant and  ${D_{\rm
LS}/D_{\rm OS}}$ is the ratio of the angular distance between the string and
the source-plane to the one between the observer and the source plane in the
thin lens  approximation.  The  projected energy density  is located  on the
intersection between the  string world-sheet and the past  light cone of the
observer  (e.g.   Fig.   \ref{intersec}).   Its  amplitude  is  given  by  a
combination of  the projected $T_{00}$, $T_{0z}$ and  $T_{zz}$ components of
the stress-energy tensor of the string if the line-of-sight is along the $z$
direction.  The displacement field is given by,
\begin{equation}
\xi_i=-\partial_i\varphi(x,y),  
\end{equation}  
and the elements of the deformation matrix can be written as,
\begin{eqnarray} 
\gamma_1&=&\left(\partial_x^2-\partial_y^2\right)\ \varphi(x,y),\\
\gamma_2&=&2\,\partial_x\partial_y\ \varphi(x,y),  
\end{eqnarray}        
the local  convergence being zero except  on the string  itself. This result
holds despite the fact that not only scalar fluctuations are contributing to
the lens effects, but also vector and tensor modes~\cite{UB}.

\section{A simplified model: the Poisson strings}

In  order to  have analytical  insights  into the  string lens  phenomelogy,
realistic  moels  for  string  shape  and energy  should  incorpate  complex
features: the  string are  very far from  being straight lines  with uniform
energy distribution.  We choose~\cite{BU} to describe the energy fluctuation
in a  simple manner, assuming that  the string follows a  straight line, but
with local  energy fluctuations.  This  fluctuations are assumed  to account
for  the various  changes  of shape,  density  of the  strings, to  possible
non-standard equation of  states, or to the existence  of currents along the
string.   We  therefore assume  the  string to  be  straight  along the  $y$
direction, and the local projected lineic energy density {\em $\mu(s)$ to be
a  random field}.   To  specify our  model  we still  need  to explicit  the
statistical properties of the $\mu$  field. Its average value obviously does
not vanish and is given by
\begin{equation}
\mg\mu(s)\md=\mu_0\equiv {\xi_0\over 4\pi G\,D_{LS}/D_{OS}},
\end{equation}
where  $\xi_0$ is  the typical  angular displacement  induced by  the string
($2\xi_0$ would be the pair separation if the string were uniform).

With the lack  of any well understanding of  the string microscopic physics,
we simply assume  the coherence length of $\mu$ to be  much smaller than the
other  distances intervening  in  the  problem.  In  this  case the  2-point
correlation function of $\mu$ can be written,
\begin{equation}
\mg\mu(s_1)\mu(s_2)\md=s_0\,\mu_0^2\,\delta_{\rm Dirac}(s_1-s_2).
\label{mucorr}
\end{equation}
An important consequence of this assumption is that, at finite distance from
the  string, the  deformation matrix  elements  are sourced  by an  infinite
amount  of independent  portions of  the  string.  They  thus obey  Gaussian
statistics. The lens  properties of the string are  thus entirely determined
by (\ref{mucorr}), independently of  what could be the one-point probability
distribution function of $\mu$.

\subsubsection{Elementary phenomenology of "Poisson string"}
\label{ElPhenomenology}

\begin{figure}
\centering {\resizebox*{9cm}{!}{\includegraphics{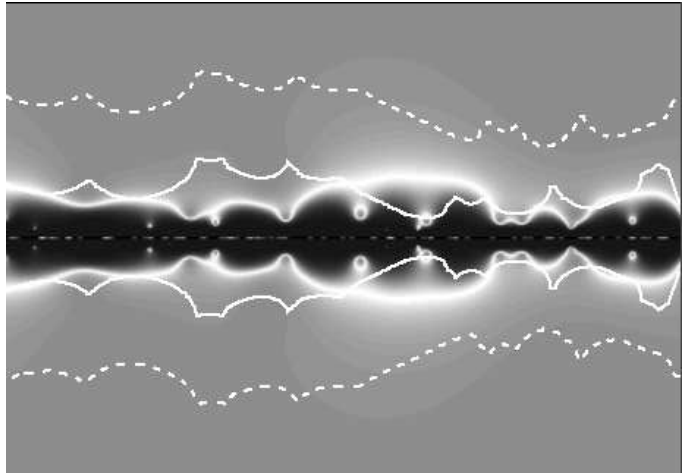}}}
\vspace{3mm}
\caption
{Numerical experiment showing the  amplification map, i.e. $\det(\mA)$, of a
``Poisson   string''.    The  brightest   pixels   correspond  to   infinite
magnification: they form the  critical lines.  The darkest pixels correspond
to  a  magnification close  to  zero.  The  solid  lines  correspond to  the
caustics, positions of the critical  lines in the source plane. The external
dashed lines are the counter images of the critical lines.}
\label{AmpMap}
\end{figure}

On Fig.   \ref{AmpMap} we depict  an example of numerical  implementation of
such a  cosmic string,  showing various features  associated with  this lens
system.  The  grey levels show the  variation of amplification  given by the
determinant  of  amplification  matrix.    Along  the  brightest  areas  the
amplification is infinite; these locations  form the critical lines. This is
where the most dramatic lens  effects could be detected: giant arcs, merging
of images...   Note that  such lines  simply do not  exist for  strings with
uniform  density!  The  critical lines  are made  mainly of  two  long lines
running along the string without crossing it.

It is worth noting  that we are here generically in a  regime of 3 images in
the vicinity of the string (except in  rare cases where it can be 5) instead
of 2 as  for a strictly uniform string. The dashed  lines show the locations
of the counter images of the critical lines.  It delimits the region, in the
image plane,  within which multiple  images can be  found.  It can  be noted
however  that the  amplification rapidly  decreases in  the vicinity  of the
string, so that  central images (i.e.  the ones situated  in between the two
infinite critical  lines) are expected  to be strongly  de-amplified, except
when they are close to one of the critical lines.

\subsection{Statistical properties}

\begin{figure}
\centering {\resizebox*{7.5cm}{!}{\includegraphics{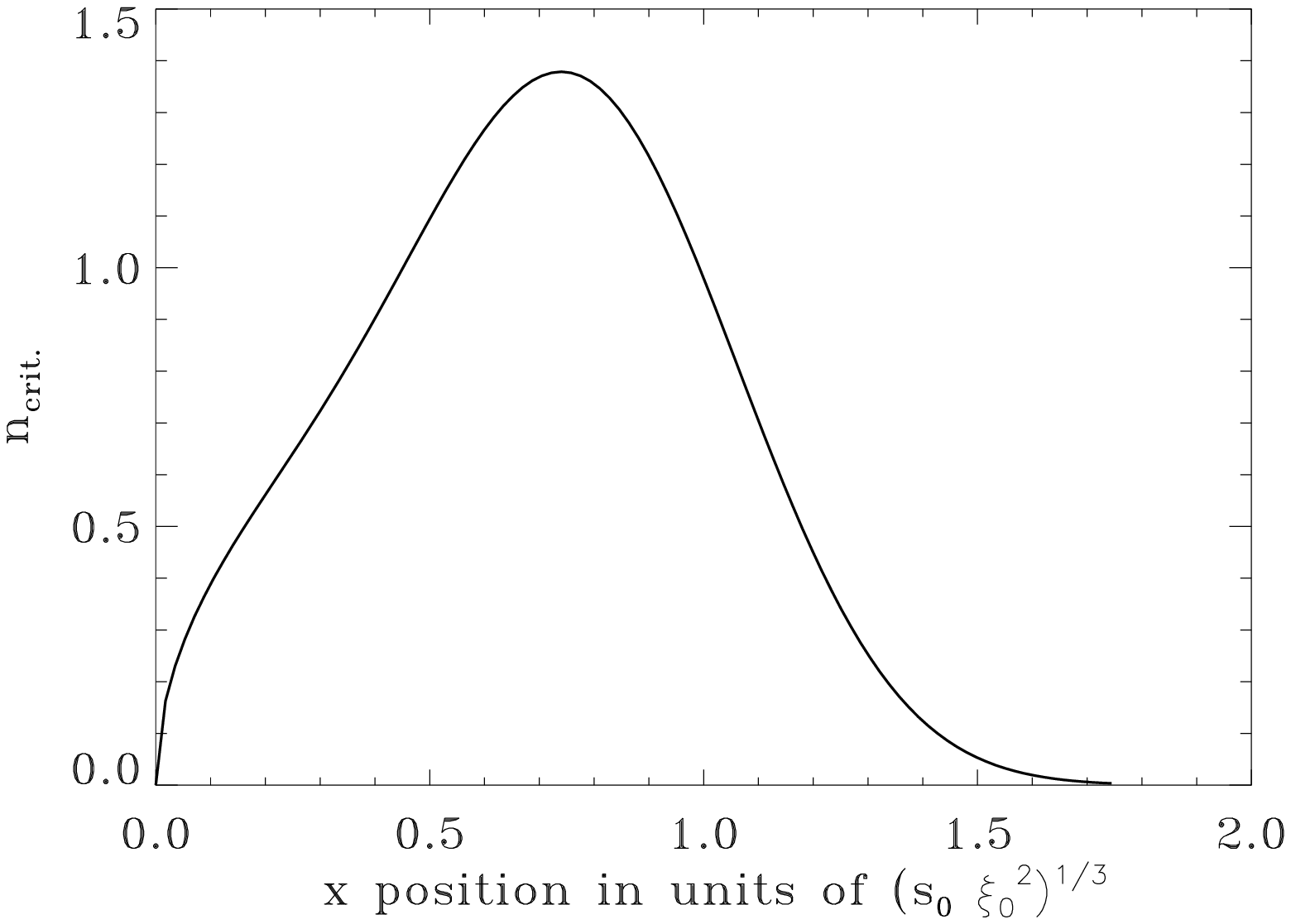}}}
\caption
{Shape of the function $\nbcrit$.}
\label{ncrit}
\end{figure}

The  necessary  conciseness of  these  notes does  not  allow  to present  a
detailed presentation of  the statistical properties of such  a system.  The
simplicity of the  model allows however the computation of  a lot of general
properties, from the position of the  critical lines, to the total length of
the caustics~\cite{BU}.

In  Fig.   \ref{ncrit}  we  present  for  instance  the  number  density  of
intersection points  of the  critical lines with  any horizontal  axis. From
this one  can infer for instance  the average number  of intersection points
(on one side of string):
\begin{equation}
  \nb=\int_0^{+\infty}\d x\,\ncrit={2\over  \sqrt{3}}\approx  1.155.  
\end{equation}
The number  of intersection points  being an odd  number, it means  that the
critical line crosses  one horizontal line more than once in  at most 7\% of
the  cases. It  supports the  fact that  we are  dominated by  the  two long
critical lines located on each side  of the string.  In rare cases, however,
inner critical lines can give rise to complex multiple image systems.

The  typical distance  of  the critical  lines  to the  string  can also  be
computed.  It is given by
\begin{equation}
\critb={\int_0^{+\infty}\d x\,x\,\nbcrit\over
\int_0^{+\infty}\d x\,\nbcrit}
\approx 0.70\,(s_0\,\xi_0^2)^{1/3}
\end{equation}
whereas the scatter of this distance is about
\begin{equation}
\Delta\critb=0.31\,(s_0\,\xi_0^2)^{1/3}.
\end{equation}

The total length of the critical lines (per unit string length) turns out to
be also calculable. Remarkably one finds
\begin{equation}
\lcrit={4\over\sqrt{3}}\,{\bf E}\left({3\over4}\right)\,
L\approx 2.80\,L,
\end{equation}
where ${\bf  E}$ is the  complete elliptic integral, {\em  independently} of
the  parameter of  the  model.  In  particular  it does  not  depend on  the
dimensionless ratio  $s_0/\xi_0$, nor on  the position of the  source plane!
And because this  result is finite, it also proves  that the closed critical
lines have a finite total length  despite the fact that they are in infinite
number.

\section{Observational prospects}

\begin{figure*}
\centering {
\resizebox*{!}{9cm}{\includegraphics{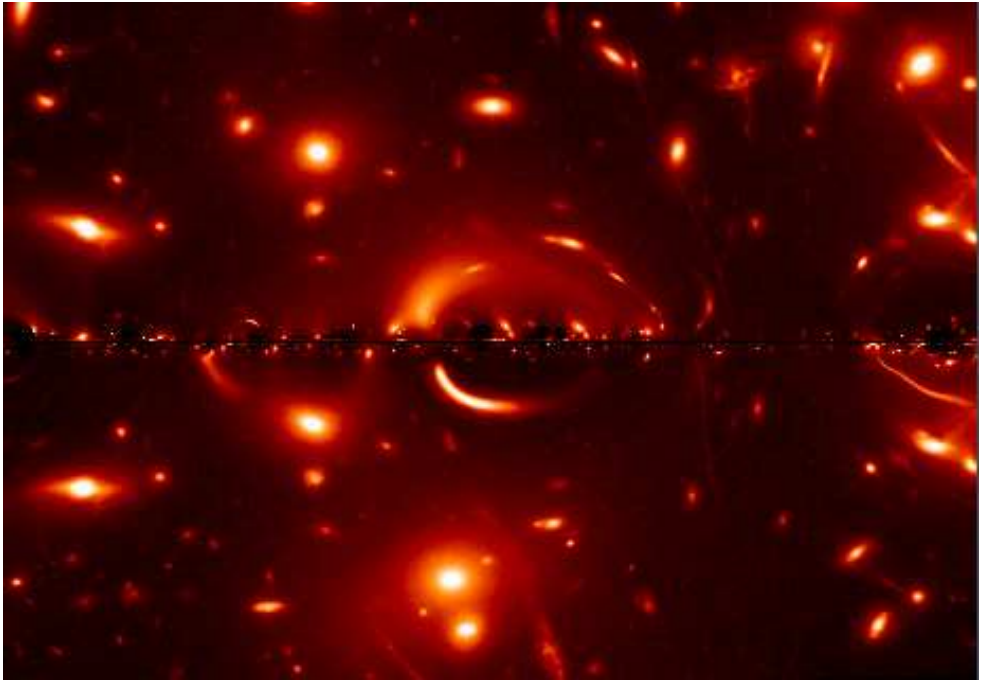}}
}
\vspace{3mm}
\caption
{Example of a deformed image by  a Poisson string.  The field corresponds to
an  external   region  of   cluster  A2218  (taken   by  the   Hubble  Space
Telescope). If  such a field was  put at $z=1$, then  an intercepting string
along the  line-of-sight would produce  multiple images as observed  on this
picture (the  typical pair separation is  about 5 arcsec)  if resolution can
reach 0.1 arcsec.  }
\label{HSTLens}
\end{figure*}

In Fig.  \ref{HSTLens}  we present a simulated image  of background galaxies
deformed by a straight Poisson string  (the case of a Poisson string loop is
depicted on  the cover).  On these  images pairs can  be clearly identified,
although distances and orientations fluctuate  from pair to pair contrary to
the standard picture.  One can also notice numerous small images that appear
along the  string. The presence of  these images are due  to the fluctuating
small scale structures of the  string.  They are associated with an infinite
number of critical lines (and  caustics) near the string. In high resolution
images, that might be the most effective way of detecting strings.

This  investigation  provides  a  new description  of  the  phenomenological
properties of lens physics for cosmic strings. The model we present is based
on general  results which state  that the lens  effects of string  are those
obtained from  lineic energy density.   The model we propose  should capture
most of  the generic properties  expected in such  a case. Although  we have
obviously not demonstrated  the validity of the description  we adopted, its
resemblance with  previous numerical results obtained  with simulated string
networks~\cite{dLKV}  makes  us think  that  this  model  could serve  as  a
guideline for detection strategies in future large angular surveys.

\section*{Acknowledgments}

We would like  to thank Y. Mellier and P. Peter  for fruitful discussions on
this subject, and T. Vachaspati for his encouragements.

\end{document}